%% file: scalar_boltzmann.tex
\documentstyle[prd,aps,12pt]{revtex}
\begin{document}

\title{The Boltzmann Equation in Scalar Field Theory}

\author{F. T. Brandt \and  J. Frenkel \and A. Guerra}

\address{\it Instituto de F\'\i sica, Universidade de S\~ao Paulo,
CP 66318, 05315-970, S\~ao Paulo, SP, Brasil}

\date{\today}

\maketitle

\vskip 1.0cm

\begin{abstract}
We derive the classical transport equation, in scalar field theory with a 
$ g^2V(\phi) $ interaction, from the equation of motion for the quantum 
field. We obtain a very simple, but iterative, expression for the effective 
action $ \Gamma $ which generates all the $ n $-point Green functions in the 
high-temperature limit. An explicit and closed form is given for $ \Gamma 
$ in the static case.
\end{abstract}

%\pacs{}

\section{Introduction}

In thermal field theory, one is often interested in the behavior of Green 
functions at high temperatures. These hard thermal functions have been much 
studied in QCD \cite{1,2,3,4,5} where they have a fundamental role
in the resummation procedure of Braaten and Pisarski \cite{6}. This 
program is usually based on quantum field theory, which involves in general 
rather complicated calculations. On the other hand, it was shown that the 
hard thermal effects can be described in a simpler and more transparent way 
using classical transport equations \cite{7,8,9}.

One may ask about the relevance of considering this approach in the
framework of a scalar field theory, given that it has been already 
applied to more realistic theories like QED and QCD. Our motivation
is twofold. First, we recall that an essential ingredient in the
derivation of classical transport equations is the equation of motion
for a particle propagating in a external background field. In QED, the
equation of motion for the electron is governed by the well known 
Lorentz force. The natural generalization to a non-Abelian gauge theory 
like QCD, is the Wong equation \cite{10a}. The geodesic equation
in gravity \cite{10} is another example where the particle has a well defined
equation of motion. The common feature of  these theories is that
they are based on a {\it gauge invariance principle} which prescribes
in a well defined way how the particle couples to the external field.
Nonlinear scalar theories, on the other hand, have no such a
simple prescription and require a rather different approach
[for an alternative treatment, see also reference \cite{10b}].
Our second purpose is to obtain an exact non-perturbative expression 
for the effective thermal action, which
contains the dominant static contributions arising in each order
of perturbation theory.
(Such a closed form result for the corresponding action in gauge theories
is generally not known.)

In section 2 we derive the Boltzmann equation, by considering the expectation 
values of the relevant quantum-mechanical quantities. The analysis shows that 
the condition for the classical limit requires that the distances over which 
the background field varies must be large compared with the particle de 
Broglie wavelength, i.e:

\begin{equation}
\frac{1}{k} \gg \frac{\hbar}{p}\label{lim}\,,
\end{equation}

\noindent
where $ k $ is a typical wavenumber of the background field and $ p $ is the 
kinetic momentum of the particle.

The behavior of Green functions at high temperatures is governed by the region involving large values of momenta of the particle propagating in the thermal loops, a domain which is consistent with the condition given by Eq. (\ref{lim}). Using the properties of the classical transport equation, we show in section 3 that the effective action which generates the high-temperature contributions of the Green functions in this domain has a form such that:

\begin{equation}
-\frac{\delta\Gamma}{\delta V^{(2)}(\phi)} = C_{(n)} \sum_{l\,=\,0}^{\infty} g^{2(l+1)} \int d^n p \left[ \frac{1}{2p\cdot\partial}\,\partial^{\mu}V^{(2)}(\phi)\,\frac{\partial}{\partial p^{\mu}}\right]^l\,f_0(p)\,,
\end{equation}

\noindent
where $ C_{(n)} $ is an appropriate normalization constant which depends on the space-time dimension $ n $, and $ f_0(p) $ is a function proportional to the Bose-Einstein probability distribution. This result is not obvious, so far as we know, since in scalar field theory the $ n $-point Green functions are sub-leading for $ n > 2 $.
 
The above action has two interesting features:

\begin{itemize}

\item a non-locality in configuration space in the form of powers of the operator $ (p\cdot\partial)^{-1} $.

\item in the static case the effective action can be expressed in closed form in terms of modified Bessel functions.

\end{itemize}

\section{Derivation of the Boltzmann Equation}

\indent

Consider the scalar field theory described by the Lagrangian density:

\begin{equation}
{\cal L} = - \frac{1}{2}\,(\partial_{\mu}\Phi)^2 - \frac{1}{2}\,\frac{m^2\, c^2}{\hbar^2}\,\Phi^2 - \frac{g^2}{\hbar^2}\,V(\Phi)\,,
\end{equation}

\noindent
where $ m $ is the mass of the field and $ V(\Phi) $ describes the interactions between the scalar fields. Examples of particular interest are the $ \Phi^4 $ theory in 4 space-time dimensions and the $ \Phi^3 $ theory in 6 dimensions. In what follows we shall work in units where $ c = 1 $, but will keep explicit the dependence on the Planck constant.

We now divide $ \Phi $ into a sum of a quantum field $ \psi $, associated with the particle, and a classical background field $ \phi $:

\begin{equation}
\Phi = \psi + \phi \,.
\end{equation}

For our purpose, it is sufficient to consider only the terms in $ {\cal L } $ which are quadratic in $ \psi $:

\begin{equation}
{\cal L}_{2} = - \frac{1}{2}\,(\partial_{\mu}\psi)^2 
-\frac{m^2\, c^2}{2\hbar^2}\,\psi^2 -\frac{g^2}{2\hbar^2}\,V^{(2)}(\phi)
\psi^2 \,.\label{L2}
\end{equation}

This leads to the linearized Euler-Lagrange equations:

\begin{equation}
\hbar^2\partial^{2}\psi = [m^2 + g^{2}V^{(2)}(\phi)]\psi\,.\label{eqmov}
\end{equation}

It is convenient to write this equation in Hamiltonian form. To this end we introduce two wave 
functions $ \psi_1 $ and $ \psi_2 $ \cite{11}

\begin{equation}
\psi = \psi_1 + \psi_2 \;\;\;,\;\;\; m\,(\psi_1 - \psi_2) = i\hbar\,\frac{\partial\psi}{\partial t}\,,
\end{equation}

\noindent
together with the single-column matrix $ \Psi $ and the Pauli matrices given by:

\begin{equation}
\Psi = \left(\begin{array}{clcr}
      \psi_1 \\
      \psi_2 \\
\end{array}\right)\, ;\label{Psi}
\end{equation}

\begin{equation}
\sigma_2 = \left(\begin{array}{clcr}
    0 & -i \\
    i &  \;\;\,0 \\
\end{array}\right)\;\;\;\;\;,\;\;\;\;\;\sigma_3 = \left(\begin{array}{clcr}
    1 &  \;\;\,0 \\
    0 & -1 \\
\end{array}\right)\, .
\end{equation}

It is then straightforward to verify that (\ref{eqmov}) is completely 
equivalent to the equation of motion:

\begin{equation}
i\hbar\,\frac{\partial\Psi}{\partial t} = \hat{H}\Psi \, ,\label{scho}
\end{equation}

\noindent
where the Hamiltonian operator $ \hat{H} $ is given by:

\begin{equation}
\hat{H} = \frac{1}{2m}\,(\sigma_3 + i\sigma_2)[\hat{\mathbf{P}}^2 + g^2V^{(2)}(\phi)] + m\sigma_3 \, ,\label{hamiltoniana}
\end{equation}

\noindent
and $ \hat{\mathbf{P}} $ is the canonical momentum operator $ -i\hbar\mathbf{\nabla} $.

In the Heisenberg representation, we have the operator equation:

\begin{equation}
i\hbar\,\frac{d\hat{{\mathbf P}}}{dt} = [\hat{{\mathbf P}},\,\hat{H}]\,.
\end{equation}

Using the form of the Hamiltonian given by (\ref{hamiltoniana}), this leads to the equation of motion for the canonical momentum:

\begin{equation}
\frac{d\hat{{\mathbf P}}}{dt} = - \frac{\,g^2}{2m}\,(\sigma_3 + i\sigma_2)
{\mathbf \nabla} V^{(2)}(\phi)\label{eqmovp}\,.
\end{equation}

We now consider the expectation value of this quantity:

\begin{equation}
<\frac{d\hat{{ \mathbf P}}}{dt}>\, =  \int 
d^{n-1}x \,\Psi^\dagger\sigma_3\,\frac{d\hat{{\mathbf P}}}{dt}\Psi\,.
\end{equation}

To evaluate the space integral, we make the assumption that the background field is roughly constant over a distance of the order of the particle de Broglie wavelength. Taking into account the spatial Lorentz contraction, we then obtain:

\begin{equation}
<\frac{d\hat{{\mathbf P}}}{d\tau}>\,= -\frac{\,g^2}{2m}\,{\mathbf \nabla} 
V^{(2)}(\phi)\label{valormedio}\,,
\end{equation}

\noindent
where $ \tau $ is the proper time of the particle.

Let us compare (\ref{valormedio}) with the classical equation of motion of a scalar particle moving in an external field, which is obtained varying the action:

\begin{equation}
{\cal S} = -\int_{\tau_i}^{\tau_f} d\tau\left[m + \frac{\,g^2}{2m}\,V^{(2)}(\phi)\right]\,.
\end{equation} 

Proceeding as usual, we get the following equation of motion:

\begin{equation}
\left[1 +\frac{\,g^2}{2m^2}\,V^{(2)}(\phi)\right]\frac{dp^\mu}{d\tau} =-\frac{\,g^2}{2m}\left[\partial^\mu V^{(2)}(\phi) - \frac{p^\mu}{p^2}\,p\cdot\partial\,V^{(2)}(\phi)\right]\,,\label{eqclassica}
\end{equation}

\noindent
where $ p_{\mu} $ denotes the kinetic momentum of the particle. Note that (\ref{eqclassica}) is consistent with the on-shell condition $ p^2 = - m^2 $ which should hold for a physical particle.

In order to compare this relation with equation (\ref{valormedio}), we replace $ p\cdot\partial\,V^{(2)} $ by $ m\,dV^{(2)}/d\tau $, which follows by the use of the chain rule. Then, we can write (\ref{eqclassica}) in the form:
 
\begin{equation}
\frac{d}{d\tau}\left\{\left[1 + \frac{\,g^2}{2m^2}\,V^{(2)}(\phi)\right]p^\mu\right\} = -\frac{\,g^2}{2m}\,\partial^\mu V^{(2)}(\phi)\,.
\end{equation}

This expression is in agreement with (\ref{valormedio}), provided the 
canonical and kinetic momentum are related by:

\begin{equation}
{\mathbf P} = \left[1 + \frac{\,g^2}{2m^2}\,V^{(2)}(\phi)\right]{\mathbf p}\,.
\end{equation}
 
Furthermore, the classical limit requires the condition:

\begin{equation}
p \gg \hbar k \,,
\end{equation}

\noindent
where $ k $ is a typical wavenumber of the background field.

Equation (\ref{eqclassica}) can be used to determine the evolution of the distribution function $ f(x,\,p) $, which represents 
the probability of finding a particle in the state $ (x,\,p) $. From the property:

\begin{equation}
\frac{df(x,\,p)}{d\tau} = 0\label{eqtransporte}\,,
\end{equation}

\noindent
which holds in the collisionless case, and using the constitutive relation \linebreak $ p_{\mu} = m\,dx_{\mu}/d\tau $, we obtain the Boltzmann equation:

\begin{equation}
p\cdot\partial\,f(x,\,p) \,-\, \frac{\,g^2}{2}\,\frac{1}{1 - g^2V^{(2)}/2p^2}\left[\partial^\mu V^{(2)} - \frac{\,p^\mu}{p^2}\,p\cdot\partial\, V^{(2)}\right]\frac{\partial f(x,\,p)}{\partial p^\mu} = 0\,.\label{eqevolu}
\end{equation}

We remark that the Planck constant does not appear in this equation, as one would have expected for a classical transport equation.

Let us expand $ f(x,\,p) $ in powers of the coupling constant:

\begin{equation}
f(x,\,p) = f_0(p) + g^2f_1(x,\,p) + g^4f_2(x,\,p) + \cdots \,,
\end{equation}

\noindent
where:

\begin{equation}
f_0(p) = \theta(p_0)\delta(p^2 + m^2)N(p_0)\,.
\end{equation}

Here, the theta and delta functions guarantee positivity of the energy and on-shell evolution, and $ N(p_0) $ is the Bose probability distribution:

\begin{equation}
N(p_0) = \frac{1}{\mathrm{e}^{\beta p_0} - 1}\, .
\end{equation}
 
Inserting the expansion of $ f(x,\,p) $ into the transport equation (\ref{eqevolu}) and identifying the coefficients of the powers of the coupling constant, we can determine recursively the components $ f_i(x,\,p) $. We find, for example:

\begin{equation}
f_1(x,\,p) = \frac{1}{2}\left[\frac{1}{p\cdot\partial}\,\partial^\mu V^{(2)}- \frac{p^\mu}{p^2}\,V^{(2)}\right]\frac{\partial f_0}{\partial p^\mu}\,,
\end{equation}

\begin{eqnarray}
f_2(x,\,p)&=&\frac{1}{4}\left[\frac{1}{p\cdot\partial}\,\partial^\mu
  V^{(2)}\,\frac{\partial}{\partial p^\mu}\right]^2 f_0(p)\nonumber \\                   &-&\frac{1}{4p^2}\,\frac{1}{p\cdot\partial}\left[\partial^\mu V^{(2)}\,\frac{\partial}{\partial p^\mu}\,(p^\nu V^{(2)})\right]\frac{\partial f_0(p)}{\partial p^\nu} + \cdots\,,
\end{eqnarray}

\noindent
where $ \cdots $ denote similar terms involving powers of $ 1/p^2 $. It turns out that such terms are not connected with the contributions obtained, in the high-temperature limit, from the 1-particle irreducible Green functions.

\section{The Effective Action}

In order to determine the form of the effective action, we will consider only those terms in the expansion of $ f(x,\,p) $ which involve powers of the operator:

\begin{equation}
D(\phi) = \frac{1}{2}\,\frac{1}{p\cdot\partial}\,\partial^{\mu}V^{(2)}(\phi)\,\frac{\partial}{\partial p^{\mu}}\,,
\end{equation} 

\noindent
which has in configuration space a non-locality of the form $ (p\cdot\partial)^{-1} $. Note that in momentum space, $ D $ is a function of degree zero in $ k $.

The above terms are generated by the series:

\begin{equation}
F(x,\,p) = \sum_{l\,=\,0}^{\infty}g^{2l}D^l(\phi)f_0(p)\,,
\end{equation}

\noindent
which satisfies the differential equation:

\begin{equation}
p\cdot\partial\,F(x,\,p) - \frac{g^2}{2}\,\partial^{\mu}V^{(2)}(\phi)\,\frac{\partial F(x,\,p)}{\partial p^{\mu}} = 0\label{eqdif}\,.
\end{equation}

It is next shown that this function is relevant in connection with the generating functional $ \Gamma $ defined by:
        
\begin{eqnarray}        
-\frac{\delta \Gamma}{\delta V^{(2)}(\phi)}& = &C_{(n)} g^2 \int d^n p \,F(x,\,p)\nonumber\\
 & = & C_{(n)} g^2 \sum_{l\,=\,0}^{\infty} g^{2l} \int d^n p\, D^l(\phi)f_0(p)\label{funcional}\, ,
\end{eqnarray}
where $C_{(n)}$ is a normalization factor.

To illustrate the content of the above quantity, let us consider for 
definiteness the $ \phi^4 $ theory in 4 space-time dimensions. 
For $ l = 0 $, we then get in momentum space the expression:

\begin{equation}
-\frac{\delta\Gamma}{\delta\phi^2} = \frac{C_{(4)}g^2}{2}\int d^4 p\, \theta(p_0)\delta(p^2 + m^2)N(p_0)\,,
\end{equation}

\noindent
which represents the scalar self-energy function $\Pi_2$ shown in Fig. 1,
provided we choose $C_{(4)}=1/(2\pi)^3$.
As is well known \cite{12} this function has a leading behavior 
proportional to $ T^2 $.

For $ l = 1 $, a simple calculation which makes use of integration by parts, gives in momentum space a contribution like:

\begin{equation}
\frac{\delta^2\Gamma}{\delta\phi^2\delta\phi^2} =
-\frac{g^4}{4\,(2\pi)^3}
\int d^4p\,\theta(p_0)\delta(p^2 + m^2)N(p_0)\,\frac{k^2}{2(p\cdot k)^2} \label{ampliclassica}\,,
\end{equation}

\noindent
where $ \hbar k $ denotes the total momentum of a pair of fields $ \phi^2 $.

We shall now compare this result with the one obtained from the 1-particle irreducible Green function shown in Fig. 2. This contribution can be expressed in terms of a forward scattering amplitude of an on-shell particle in an external field 
\cite{5} as indicated by the two graphs in Fig. 3. Also, an integration over 
the particle's momenta $ p^\mu $ must be performed with a weight factor given by the Bose probability distribution $ N(p_0) $. We then obtain a contribution of the form:

\begin{equation}
\Pi_4 = \frac{g^4}{4} \int \frac{d^3p}{(2\pi)^3}\,\frac{N(\omega)}{2\omega}\left[\frac{1}{\hbar^2k^2 + 2\hbar k\cdot p} + \frac{1}{\hbar^2k^2 - 2\hbar k\cdot p}\right]\label{quatropontos}\,.
\end{equation}

We now consider the high-temperature behavior, which is governed by the region involving high values of $ p $. Expanding the Feynman propagators in this region, we get:

\begin{equation} 
\frac{1}{\hbar^2k^2 + 2\hbar k\cdot p} + \frac{1}{\hbar^2k^2 - 2\hbar k\cdot p} = -\frac{k^2}{2(p\cdot k)^2} + {\cal O}\,(\frac{\,\hbar^2k^2}{p^4})\,.
\end{equation}

\vspace{2 mm}

Note that the leading term, which is of degree zero in $ k $, is classical since the $ \hbar $ dependence cancels out. Therefore, as far as the high-temperature domain is concerned, (\ref{quatropontos}) effectively reduces to the classical amplitude (\ref{ampliclassica}), yielding a $ \ln T $ contribution.

These calculations show that the operator $ D $ generates iteratively, in the high-temperature limit, the dominant contributions associated with the 1-particle irreducible Green functions. It is clear that this mechanism generalizes to more general forms of interactions between the scalar fields.
Hence, the generating functional $ \Gamma $ given by equation (\ref{funcional}) can be interpreted as the high-temperature effective action in scalar field theory.

We have not been able to evaluate $ \Gamma $ in closed form for arbitrary external fields. However, this is possible when the fields are static, in which case the function $ F(x,\,p) $ can be determined exactly. It is easy to verify that the static solution of the equation (\ref{eqdif}), which reduces to $ f_0 $ in the absence of interactions, is given by:

\begin{equation}
F(x,\,p) = \theta(p_0)\delta(p^2 + m^2 + g^2V^{(2)}(\phi))N(p_0)\,.
\end{equation}

With this form, (\ref{funcional}) may be integrated to give the following local expression:

\begin{equation}
\Gamma = C_{(n)} \int  d^nx \int  d^np\,\theta(p_0)\theta[-(p^2 + m^2 + g^2V^{(2)}(\phi))]N(p_0)\,.
\end{equation}

Using the constraints imposed by the $ \theta $ functions, the $ {\mathbf p} $ integrations can be done, with the result:

\begin{equation}
\Gamma = \frac{2C_{(n)}}{n - 1}\,\frac{\pi^{(n - 1)/2}}{\Gamma[(n - 1)/2]}\int  d^nx \int_M^{\infty}  dp_0\,({p_0}^{2} - M^2)^{(n - 1)/2}N(p_0)\,,
\end{equation}

\noindent
where $ M^2 = m^2 + g^2V^{(2)}(\phi) $.

After expanding $ N(p_0) $ as a power series in exponentials, the $ p_0 $ integrations can be expressed in terms of modified 
Bessel functions \cite{13}. We then obtain, in the static case, the following expression for $ \Gamma $:

\begin{equation}
\Gamma = \pi^{n/2 - 1}C_{(n)}\sum_{l\,=\,1}^{\infty}\left(\frac{2T}{l}\right)^{n/2}  \int  d^nx\, M^{n/2}K_{n/2}\left(\frac{lM}{T}\right)\,.\label{Bessel}
\end{equation}
 
The properties of this result may be illustrated in the case of the $ \phi^4 $ theory in 4 space-time dimensions. Making a high-temperature expansion of the Bessel function $ K_2 $ in powers of $ lM/T $, and dropping an irrelevant constant term, we obtain a series of the form:

\begin{equation}
\Gamma = -\frac{T^2}{8\pi^2} \int  d^4x\,M^2\left[\frac{\,\pi^2}{3} +\frac{1}{8}\frac{M^2}{T^2}\ln\frac{M^2}{T^2} + \frac{\xi^{(1)}(-2)M^4}{24T^4} + \cdots\right]\,,\label{final}
\end{equation}

\noindent
where $ \xi^{(1)} $ is the derivative of the Riemann zeta function and
\linebreak $ M^2 = m^2(1 + g^2\phi^2/2m^2) $.

The first term in equation (\ref{final}) corresponds to the leading $
T^2 $ contribution associated with the scalar self-energy
function. The second term contains, in addition to sub-leading
self-energy function, the $ \ln T $ contribution of the 4-point
function in the high-temperature limit. The contributions associated
with the scalar 6-point function first appear in the third term.

It is interesting to remark that the expression \ref{final} also contains
a non-perturbative term involving $ \ln [1 + g^2\phi^2/2m^2] $. 
This contribution may be expanded in a power series of the coupling 
constant when $g^2\phi^2 < 2m^2 $, but in very strong external fields the
perturbative expansion is no longer meaningful.
Hence, aside from the static condition, we expect that
the result (\ref{Bessel}) may be applicable in perturbation
theory for not too strong external fields.
 
\vfill\eject

\pagebreak

\section*{Figure Captions}

\indent

Fig. 1 The scalar self-energy function $\Pi_2$.

Fig. 2 The irreducible 4-point function $\Pi_4$.

Fig. 3 The forward scattering amplitudes associated with the 4-point function.

\vspace{20 mm}

\input{f2artigo}

\vspace{8 mm}

\input{f4artigo}

\vspace{8 mm}

\input{fqartigo}

\end{document}

%% file: f2artigo.tex
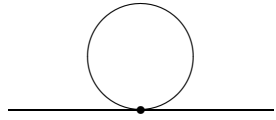
\begin{figure}[h,b]

\centering

\begin{picture}(0,30)

\put(-50,0){\line(2,0){100}}

\put(0,20){\circle{40}}

\put(0,0){\circle*{3}}

\end{picture}

\caption{}

\end{figure}

%% file: f4artigo.tex
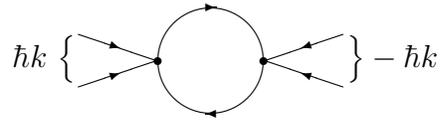
\begin{figure}[h,t,b]

\centering

\begin{picture}(0,30)

\put(0,20){\circle{40}}

\put(-20,20){\circle*{3}}

\put(20,20){\circle*{3}}

\put(-50,30){\line(3,-1){30}}

\put(-35,25){\vector(3,-1){1}}

\put(-50,10){\line(3,1){30}}

\put(-35,15){\vector(3,1){1}}

\put(-75,17.5){$ \hbar k \, \left\{ \frac{}{} \right. $}
                                     
\put(50,30){\line(-3,-1){30}}

\put(35,25){\vector(-3,-1){1}}

\put(50,10){\line(-3,1){30}}

\put(35,15){\vector(-3,1){1}}

\put(48,17.5){$ \left. \frac{}{} \right\} -\hbar k $}

\put(1,40){\vector(2,0){1}}

\put(-1,0){\vector(-2,0){1}}

\end{picture}

\caption{}

\end{figure}

%% file: fqartigo.tex
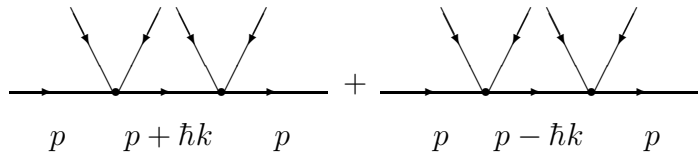
\begin{figure}[h,b,t]

\centering

\begin{picture}(0,30)

\put(-130,20){\line(1,0){120}}

\multiput(-90,20)(40,0){2}{\circle*{3}}

\multiput(-91,20)(40,0){2}{\line(-1,2){16}}

\multiput(-101,40)(40,0){2}{\vector(1,-2){1}}

\multiput(-89,20)(40,0){2}{\line(1,2){16}}

\multiput(-79,40)(40,0){2}{\vector(-1,-2){1}}

\put(-115,0){$ p $}

\put(-115,20){\vector(1,0){1}}

\put(-90,0){$ \; p + \hbar k $}

\put(-70,20){\vector(1,0){1}}

\put(-30,0){$ p $}

\put(-30,20){\vector(1,0){1}}

\put(10,20){\line(1,0){120}}

\multiput(50,20)(40,0){2}{\circle*{3}}

\multiput(49,20)(40,0){2}{\line(-1,2){16}}

\multiput(39,40)(40,0){2}{\vector(1,-2){1}}

\multiput(51,20)(40,0){2}{\line(1,2){16}} 

\multiput(61,40)(40,0){2}{\vector(-1,-2){1}}

\put(30,0){$ p $}

\put(30,20){\vector(1,0){1}}

\put(50,0){$ \; p - \hbar k $}

\put(70,20){\vector(1,0){1}}

\put(110,0){$ p $}

\put(110,20){\vector(1,0){1}}

\put(-4,20){$ + $}

\end{picture}

\caption{}

\end{figure}